\documentclass[9pt,twocolumn,twoside,lineno]{pnas-new}
\templatetype{pnasresearcharticle} % Choose template
\begin{document}
\title{Threshold Current Density for Diffusion-controlled Stability of Electrolytic Surface Nanobubbles}

% Use letters for affiliations, numbers to show equal authorship (if applicable) and to indicate the corresponding author
\author[a]{Yixin Zhang}
\author[b]{Xiaojue Zhu}
\author[c]{Jeffery A. Wood}
\author[a,d,1]{Detlef Lohse}

\affil[a]{Physics of Fluids Group, Max Planck Center Twente for Complex Fluid Dynamics and J. M. Burgers Centre for Fluid Dynamics, University of Twente, P.O. Box 217, 7500 AE Enschede, The Netherlands}
\affil[b]{Max Planck Institute for Solar System Research, Göttingen, 37077, Germany}
\affil[c]{Membrane Science and Technology Cluster, MESA+ Institute for Nanotechnology, University of Twente, the Netherlands}
\affil[d]{Max Planck Institute for Dynamics and Self-Organization, 37077 Göttingen, Germany}

% Please give the surname of the lead author for the running footer
\leadauthor{Zhang}

% Please add a significance statement to explain the relevance of your work
\significancestatement{The generation of micro- and nanobubbles on electrodes significantly impedes electrolysis efficiency by blocking electrolyte access to the electrodes. How to get rid of them and when do they detach? Progress on this question has been hindered by a limited understanding of the process of nanobubble nucleation, growth, and stability on the electrodes. Here we study this problem by performing molecular simulations of water splitting, observing either sticking bubbles or detaching bubbles, depending on the current density. Building on the stability theory for surface nanobubbles, we develop a theory with which we can successfully predict the threshold current density. Our work has important implications for enhancing bubble detachment and thus the efficiency of electrolysis.}

% Please include corresponding author, author contribution and author declaration information
\authorcontributions{Y. Z and D. L: study conception, simulations and manuscript preparation. X. Z and J. A. W.: manuscript preparation and discussions.}
\authordeclaration{The authors declare no competing interest.}
%\equalauthors{\textsuperscript{1}A.O.(Author One) contributed equally to this work with A.T. (Author Two) (remove if not applicable).}
\correspondingauthor{\textsuperscript{1}To whom correspondence should be addressed. E-mail: d.lohse@utwente.nl}

% At least three keywords are required at submission. Please provide three to five keywords, separated by the pipe symbol.

\begin{abstract}
Understanding the stability mechanism of surface micro/nanobubbles adhered to gas-evolving electrodes is essential for improving the efficiency of water electrolysis, which is known to be hindered by the bubble coverage on electrodes. Using molecular simulations, the diffusion-controlled evolution of single electrolytic nanobubbles on wettability-patterned nanoelectrodes is investigated. These nanoelectrodes feature hydrophobic islands as preferential nucleation sites and allow the growth of nanobubbles in the pinning mode. In these simulations, a threshold current density distinguishing stable nanobubbles from unstable nanobubbles is found. When the current density remains below the threshold value, nucleated nanobubbles grow to their equilibrium states, maintaining their nanoscopic size. However, for the current density above the threshold value, nanobubbles undergo unlimited growth and can eventually detach due to buoyancy. {Increasing the pinning length of nanobubbles increases the degree of nanobubble instability.} By connecting the current density with the local gas oversaturation, an extension of the stability theory for surface nanobubbles [Lohse and
Zhang, Phys. Rev. E, 2015, 91, 031003(R)] accurately predicts the nanobubble behavior found in molecular simulations, including equilibrium contact angles and the threshold current density. For larger systems that are not accessible to molecular simulations, continuum numerical simulations with the finite difference method combined with the immersed boundary method are performed, again demonstrating good agreement between numerics and theories.
\end{abstract}

\dates{This manuscript was compiled on \today}
\doi{\url{www.pnas.org/cgi/doi/10.1073/pnas.XXXXXXXXXX}}

\maketitle
\thispagestyle{firststyle}
\ifthenelse{\boolean{shortarticle}}{\ifthenelse{\boolean{singlecolumn}}{\abscontentformatted}{\abscontent}}{}

\firstpage[16]{2}
% Use \firstpage to indicate which paragraph and line will start the second page and subsequent formatting. In this example, there are a total of 11 paragraphs on the first page, counting the first level heading as a paragraph. The value {12} represents the number of the paragraph starting the second page. If a paragraph runs over onto the second page, include a bracket with the paragraph line number starting the second page, followed by the paragraph number in curly brackets, e.g. "\firstpage[4]{11}".
The most promising solution towards achieving a zero-carbon society involves electrochemical water splitting to produce hydrogen, powered by renewable electricity \cite{shih2022water,brauns2020alkaline,yue2021hydrogen}.
Hydrogen plays a ubiquitous role in our daily lives, with applications ranging from refining petroleum, fertilizer production, food processing, and plastics manufacturing to transportation \cite{ramachandran1998overview}. However, realizing the vision of a sustainable hydrogen economy necessitates a substantial scale-up of ongoing hydrogen production. Central to this effort is to increase the current density, a key parameter in electrochemical processes \cite{shih2022water,brauns2020alkaline,yue2021hydrogen}. The formation of micro and nanobubbles on gas-evolving electrodes is believed to block the active electrode area and thus increase the overpotential and decreasing the current density \cite{vogt2005bubble,angulo2020influence,zhao2019gas}. Effectively addressing this issue hinges on our comprehensive understanding of the life cycle of individual nanobubbles on electrodes, encompassing their nucleation, growth, and detachment processes. The advancement of our knowledge in hydrogen bubble evolution will also benefit many other applications where electrochemical gas evolution exists such as the chlorine evolution reaction in the chloralkaline process \cite{karlsson2016selectivity}, the hydrazine oxidation reaction in fuel cells \cite{lu2015superaerophobic} and the aerosol emitted in an electrowinning system \cite{papachristodoulou1985bubble}.

Due to the small spatial and temporal scales of nanobubbles, it remains challenging to produce single nanobubbles and observe them directly as single entities \cite{lohse2015surface}. Atomic force microscopy (AFM) has frequently been used to image the density of nanobubbles and their sizes \cite{yang2009electrolytically,wang2010boundary,zhao2013mechanical,yu2023interfacial} but this technique usually requires an electrode size of micrometers where multiple nanobubbles are generated. The group of White managed to use nanoelectrodes as an innovative method to generate single nanobubbles \cite{luo2013electrogeneration,liu2017dynamic,chen2015electrochemical,edwards2019voltammetric}. Though the formed single nanobubbles cannot be observed directly, their presence and equilibrium states are indicated by a sudden drop of the peak current to a steady value. {Single nanobubbles cannot easily be directly observed. Their presence and equilibrium states are often indicated by a sudden drop of the peak current to a steady value. More recently nanobubbles have been imaged using the off-axis dark-field microscopy \cite{suvira2023imaging}, although the resolution was still insufficient to determine the exact dimensions of the nanobubble.} Nanopipettes or nanopores are also used to
produce single nanobubbles \cite{zhou2023nanopipettes}. However, integrating nanopipettes with advanced microscopy and spectroscopy techniques is still needed to image single nanobubbles. Hao et al. \cite{hao2018imaging} used super-resolution microscopy to image transient formation and growth of single hydrogen nanobubbles based upon a single-molecule labeling process where fluorescence dye molecules adsorb onto the bubble interface. Deng et al. \cite{deng2022direct} utilized scanning electrochemical cell microscopy techniques to measure the single heterogeneous bubble nucleation on a nanoparticle. Lemineur et al. \cite{lemineur2021imaging} proposed using interference reflection microscopy to analyze the geometry and growth rate of individual nanobubbles on nanoparticles.

Complementary to experimental studies, molecular dynamics (MD) simulations can provide excellent spatial and temporal resolutions of electrolytic nanobubbles \cite{perez2019mechanisms,maheshwari2020nucleation,ma2021dynamic,gadea2020electrochemically,zhang2023minimum}. Sirkin et al. \cite{perez2019mechanisms} used molecular simulations with an algorithm that mimics the electrochemical formation of gas, to investigate the mechanisms of nucleation of gas bubbles on nanoelectrodes and characterize their stationary states. Maheshwari et al. \cite{maheshwari2020nucleation} studied the nucleation and growth of a nanobubble on rough surfaces using molecular dynamics simulations. They show that the oversaturation of gas required for nucleation of a nanobubble depends on the surface morphology. Using MD simulations, Ma et al. \cite{ma2021dynamic} showed that gas solubility or solute concentration results in various nanobubble dynamic states at a nanoelectrode such as pinned bubbles or unpinned bubbles.

Despite these experimental and numerical efforts, the current theoretical understanding of the dynamics of single electrolytic nanobubbles is still developing. In terms of the single nanobubble generated on the nanoelectrode in experiments  \cite{luo2013electrogeneration,liu2017dynamic,chen2015electrochemical,edwards2019voltammetric,suvira2023imaging}, even the simple question ‘What is the contact angle of the pinned bubble given the value of the current?’ is still surprisingly difficult to answer. When surface nanobubbles were first discovered in the 1990s \cite{parker1994bubbles}, it was also difficult to explain their features (long lifetime and small contact angles \cite{zhang2006physical,lou2000nanobubbles}) as the classic Epstein-Plesset equation \cite{epstein1950stability} predict they should dissolve in microseconds. After about two decades of progress (see again the review \cite{lohse2015surface}), the stability mechanism of surface nanobubbles is now well explained by the Lohse-Zhang model \cite{lohse2015pinning} which implies that contact line pinning and local oversaturation is necessary for the stability of surface nanobubbles. Electrolytic nanobubbles essentially belong to the family of surface nanobubbles. Very recently this Lohse-Zhang model has been generalized to electrolytic surface nanobubbles by including the gas influx produced at the contact line \cite{zhang2023minimum}, which can be used to estimate the contact angles of single electrolytic surface nanobubbles found in experiments \cite{luo2013electrogeneration,liu2017dynamic,chen2015electrochemical,edwards2019voltammetric,suvira2023imaging} and molecular simulations \cite{zhang2023minimum}.

However, previous studies \cite{luo2013electrogeneration,liu2017dynamic,chen2015electrochemical,edwards2019voltammetric,suvira2023imaging,perez2019mechanisms,suvira2023imaging,maheshwari2020nucleation,ma2021dynamic,gadea2020electrochemically,zhang2023minimum} mainly focus on the \emph{reaction-controlled} evolution of single nanobubbles where the produced gas on the electrode goes directly into the bubble. In fact, the power-law growth of bubbles with time $R\sim t^{\beta}$ ($R$ is the bubble radius of curvature) at larger scales (above micrometers) in water electrolysis is known to have two main modes depending on the values of the Damk\"{o}hler number $\text{Da}=A_e/R^2$, which is the ratio of active electrode area to the bubble surface area (typically the $\text{Da}$ number is expressed in reaction rates over diffusion which in this case reduces to the ratio defined here) \cite{van2017electrolysis,higuera2021model}. For $\text{Da} \ll 1$, i.e., relatively small active electrode surfaces, the bubble growth mode is called as the aforementioned \emph{reaction-controlled} growth and $\beta=1/3$. This is the case for a series of experiments done by the group of White \cite{luo2013electrogeneration,liu2017dynamic,edwards2019voltammetric} where the formed bubble blocks almost the entire electrode, leaving only the places of contact lines to generate gas. The generated gas goes directly into the bubble following energy minimization. Conversely, for $\text{Da} \gg 1$, i.e., relatively large active electrode surfaces, the bubble growth mode is known as the \emph{diffusion-controlled} growth and $\beta=1/2$. The produced gas diffuses into the bulk liquid and builds up the oversaturation around the bubble, which leads to the evolution of the nucleated bubble. This growth mode has been extensively studied in experiments done in large scales \cite{van2017electrolysis,wang2016investigations,verhaart1980growth}.
However, the \emph{diffusion-controlled} mode for single electrolytic surface nanobubbles at the nanoscale has been reported far less in experiments or simulations, which may be significantly different from the growth at the macroscale. In fact, since electrolytic bubbles are formed by nucleation and they are very small initially, understanding their early-stage growth is thus crucial for preventing the appearance of bubble blockage on electrodes.

Here we propose using a wettability-patterned nanoelectrode to generate single nanobubbles where a hydrophobic nano-island is positioned within a hydrophilic nanoelectrode. This will promote preferential nucleation of bubbles on the hydrophobic island, allowing a well-controlled study of the diffusion-controlled bubble growth. We perform molecular simulations to demonstrate this. {This setting also captures the scenario of preferential nucleation and pinning of nanobubbles by cavities that are inherent on nanoelectrodes \cite{lohse2016homogeneous,jones1999bubble,angulo2020influence}}. Within these simulations, we systematically vary the current density to observe its impact on the contact angles of nucleated nanobubbles and assess their stability. For a larger system not accessible to molecular simulations, a finite difference (FD) method coupled to the immersed boundary method is adopted to perform the simulations. The Lohse-Zhang model is extended to predict the bubble behaviours found in the MD and the FD simulations.

% If your first paragraph (i.e. with the \dropcap) contains a list environment (quote, quotation, theorem, definition, enumerate, itemize...), the line after the list may have some extra indentation. If this is the case, add \parshape=0 to the end of the list environment.
%\dropcap{T}his PNAS journal template is provided to help you write your work in the correct journal format. Instructions for use are provided below.

%Note: please start your introduction without including the word ``Introduction'' as a section heading (except for math articles in the Physical Sciences section); this heading is implied in the first paragraphs.

\section*{Results and Discussions}
\begin{figure*}[t!]
\centering
\includegraphics[width=17.8cm]{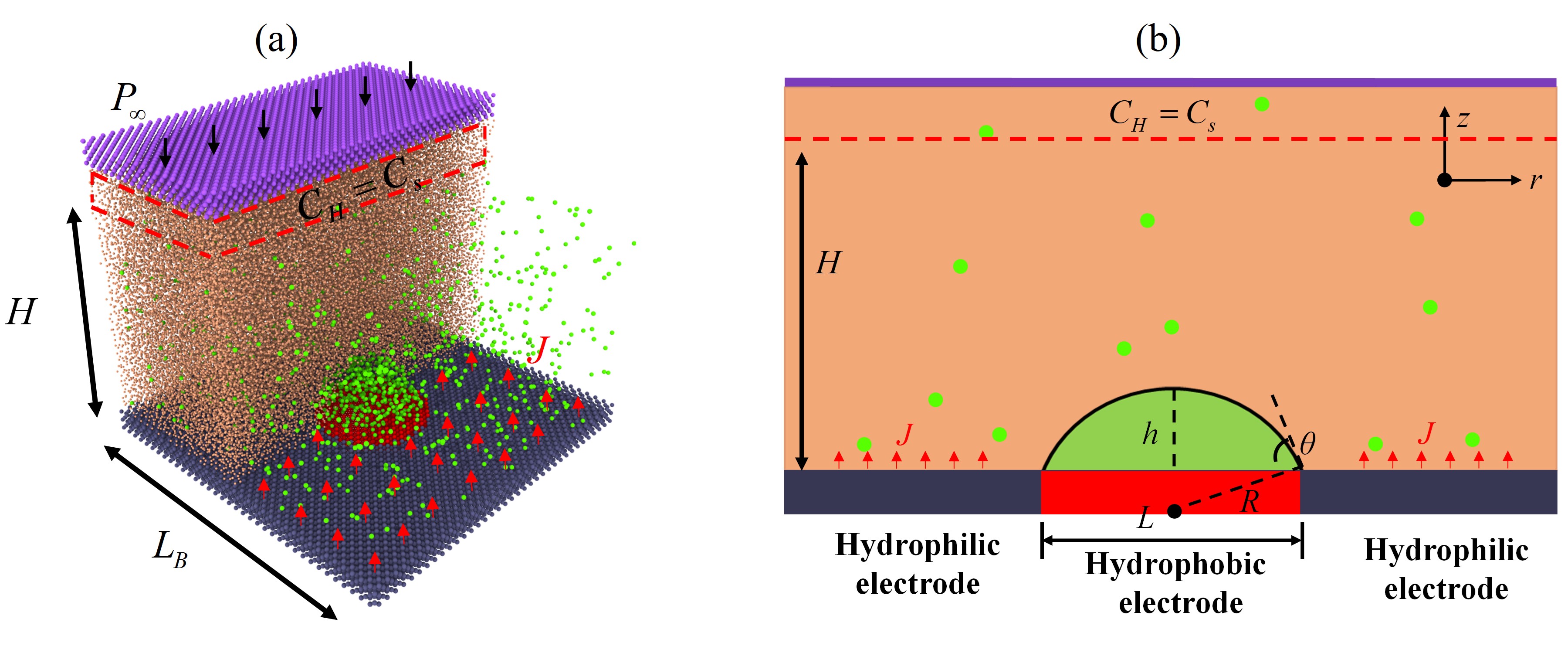}
\caption{(a) A snapshot of the generated electrolytic nanobubble on the nanoelectrode in MD simulations. The simulated domain has been sliced to observe the bubble. The system's condition is maintained at $T=300$ K, $P_{\infty}=10$ atm, and $C_{H}=C_s$. The square solid plate at the bottom with a length $L_B=17.28$ nm is the wetability-patterned nanoelectrode. The disk in the middle has a diameter $L=5.76$ nm and is made hydrophobic to water, while the surrounding solid is hydrophilic to water. The $J$ represents the gas influx produced on the electrode. For small bubbles, $H$ is nearly constant and $H=11.25$ nm. (b) Sketch of a nanobubble with pinning length $L$, contact angle $\theta\left(t\right)$, radius of curvature $R\left(t\right)$ and height $h\left(t\right)$.}  \label{fig1}
\end{figure*}

\subsection*{Molecular simulations of water electrolysis}Molecular dynamics (MD) simulations are used as virtual experiments to simulate the generation of electrolytic nanobubbles on wettability-patterned nanoelectrodes. The popular open-source code LAMMPS \cite{plimpton1995fast} is adopted. As shown in Figure \ref{fig1}(a), the minimal molecular system consists of monatomic water molecules (represented in orange color), monatomic gas molecules (represented in green color), atoms of the hydrophobic electrode represented in red color and hydrophilic electrode represented in blue color, and atoms of the `piston' plate (denoted in purple color). The system's condition is maintained at $T=300$ K using the Nos\'{e}-Hoover thermostat and $P_{\infty}=10$ atm using the piston in a standard way \cite{ma2021dynamic,dockar2018mechanical}. {The use of $P_{\infty}=10$ atm instead of $P_{\infty}=1$ atm is done in order to increase the gas solubility and thus reduce the statistic errors in such a small molecular system.}

The water molecule is modeled by the monatomic mW water potential \cite{molinero2009water,perez2019mechanisms} for the save of computational costs and the relatively good accuracy of the water surface tension $\gamma=66$ mN/m. The gas is modeled by the standard 12-6 Lennard-Jones (LJ) potential and has a density $\rho_{\infty}=11.47$ kg/m\textsuperscript{3} at 10 atm and 300 K. All used parameters are provided in detail in the section of Methods. The gas-water interaction is tuned to obtain a gas solubility $C_s=0.54$ kg/m\textsuperscript{3} and a mass diffusivity $D=4.3\times 10^{-9}$ m\textsuperscript{2}/s (see the Methods for how these values are obtained). 

The process of water splitting is modeled in a simple way as in previous MD studies \cite{perez2019mechanisms,maheshwari2020nucleation,ma2021dynamic}. Above the electrode, two layers of water molecules can turn into gas atoms (by switching atom types) conducted at a fixed rate, leading to a constant gas influx $J$ in units of kg/(m\textsuperscript{2}s), i.e., a constant current density $i_{in}=JnF/M_g$ in units of A/m\textsuperscript{2}. Here we assume the production of each gas atom needs $n=1$ electron in our simulations. $F$ is Faraday constant. $M_g=0.028$ kg/mol is the molar mass of the gas simulated in the current system. {Atoms in these two water layers have the same probability (uniform probability distribution) to be converted into gas and are randomly selected for this conversion.} Notably as in previous MD simulations \cite{perez2019mechanisms,maheshwari2020nucleation,ma2021dynamic}, we are not trying to simulate a gas which exactly represents hydrogen. Since hydrogen has a very low solubility in water, a larger system would be required to reduce statistic errors but the computational costs are not affordable. {Our simulation closely represents the experiments of water electrolysis operated at the condition of a constant current \citep{bard2022electrochemical,penas2019decoupling}. If the experiments are performed with a constant voltage, a dynamic equilibrium state of stable surface nanobubbles may be achieved where the current is constant over time \cite{yang2009electrolytically}. Previous works of surface nanobubbles, whether on nanoelectrodes or not, \citep{chen2015electrochemical,zhang2006physical,higuera2021model} show that the addition of ions or electrostatic forces on nanobubble surfaces have minor effects on nanobubble stability, as the Laplace pressure inside the bubble is very high. Ions and electrostatic forces are thus not considered in the current simulations.}

\subsection*{Stable and unstable electrolytic nanobubbles in MD simulations}
\begin{figure*}[t!]
\centering
\includegraphics[width=\linewidth]{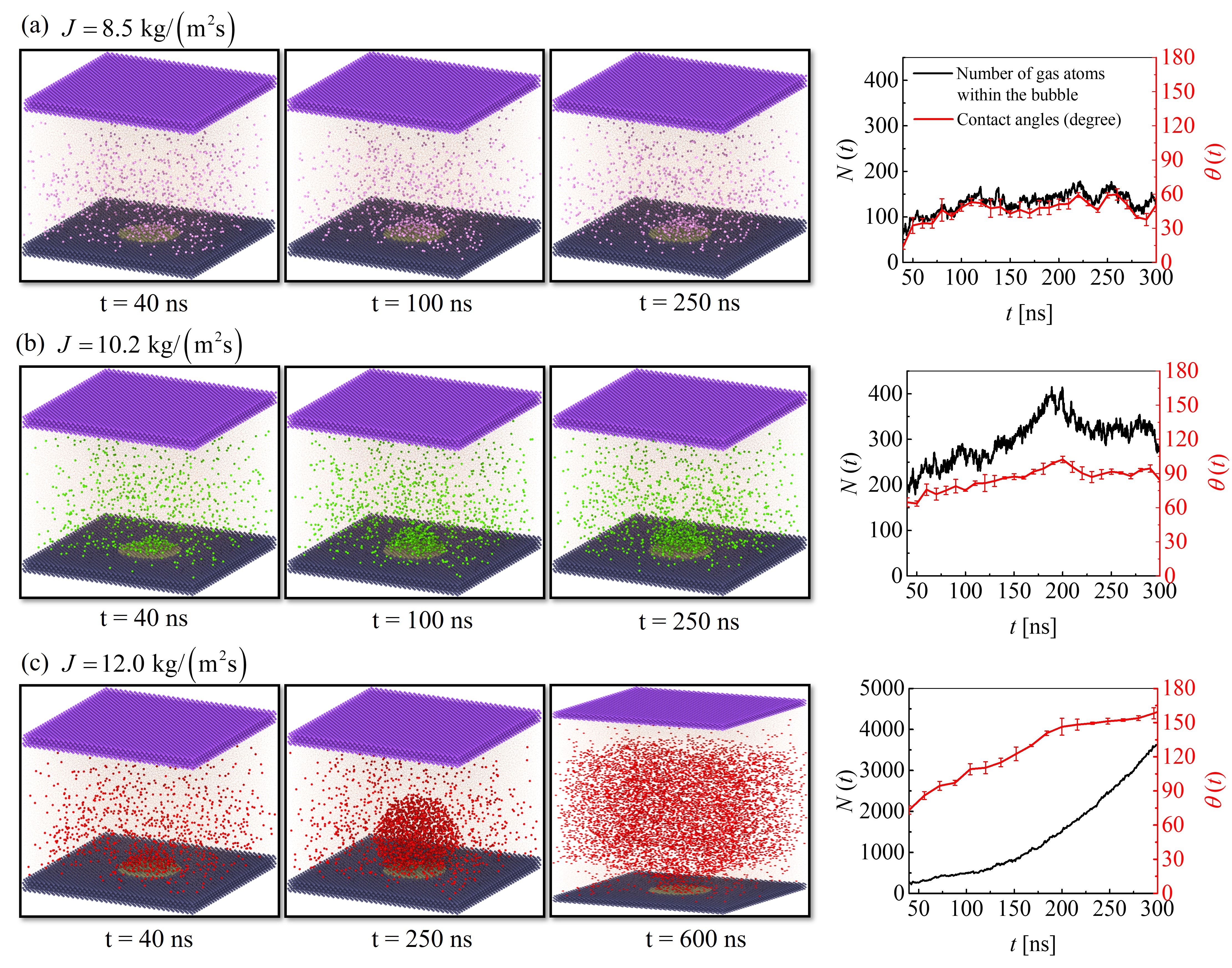}
\caption{Status of electrolytic nanobubbles under three different gas fluxes. The snapshots show the evolution of nanobubbles at three different time. The figure in the right panel shows the evolution of contact angels and the number of gas atoms. The error bar represents the standard deviation of ten measurements, see SI for more details. (a) $J=8.5$ kg/(m\textsuperscript{2}s), stable nanobubble; (b) $J=10.2$ kg/(m\textsuperscript{2}s), stable nanobubble; (c) $J=12.0$ kg/(m\textsuperscript{2}s), unstable bubble and the bubble breaks the system as shown by the snapshot at $t=600$ ns.}  \label{fig2}
\end{figure*}
In our simulations, the gas flux is varied, focusing on possible bubble nucleation at the electrode. For a small value of gas fluxes, e.g., $J=3.1$ kg/(m\textsuperscript{2}s), there is no bubble nucleation due to the low levels of oversaturation at the electrode. For a larger gas flux, bubble starts to nucleate on the hydrophobic electrode and then grows in the mode of a constant pinning length. The contact line pinning is the result from the chemical heterogeneity between the hydrophobic part of the electrode and the hydrophilic part of the electrode. {Preferential nucleation and pinning of macrobubbles achieved by partially hydrophobic electrodes have also been explored in previous experiments \citep{brussieux2011controlled}.} Figures \ref{fig2} (a) and (b) (also see Movie S1 and Movie S2 in the SI) show transient snapshots of nucleated bubbles with $J=8.5$ kg/(m\textsuperscript{2}s) and $J=10.2$ kg/(m\textsuperscript{2}s) respectively, showing that the nanobubbles experience growth at first and eventually reach their stationary states with contact angles depending on the value of gas fluxes. The evolution of contact angles and the number of gas atoms inside the bubbles for these two cases is shown in the right panel of Figures \ref{fig2} (a) and (b), demonstrating that the bubbles eventually become equilibrated. {The methodology for determining the contact angles and the number of gas atoms inside the bubble is standard and has been described in our previous work \cite{zhang2019molecular,weijs2011origin}; details are also provided in SI.} The relation between equilibrium contact angles measured from MD simulations and the gas influxes ($J=6.8, 8.5, 9.4, 10.2$ kg/(m\textsuperscript{2}s)) is shown by symbols in Figure \ref{fig3}. 

However, for even larger gas fluxes, e.g., $J=12.0$ kg/(m\textsuperscript{2}s), as shown by the snapshots in Figure \ref{fig2} (c) (also see the supplement Movie S3), {the nucleated nanobubble is unstable and grows beyond the system size as it interacts with its periodic image and separates the liquid from the electrode. The number of gas atoms inside the bubble increases very rapidly as recorded in the right panel of Figure \ref{fig2} (c), demonstrating that the bubble cannot be stable. The depletion of liquid from the electrode by the gas film may be thought of as a similar process to the transition from nucleate boiling to film boiling as the result of the coalescence of vapor bubbles \citep{rohsenow1971boiling}. However, the growth of gas bubbles in electrolysis is typically slower than the growth of vapor bubbles in boiling as the heat diffusion is about three orders of magnitude faster than mass diffusion \citep{prosperetti2017vapor}.}    

Such a transition from stable nanobubbles to unstable bubbles by increasing the current density (gas flux) is very crucial as it helps to explain when nanobubbles adhere to the electrode and how they can detach: there exists a threshold current density $i_t$ above which nucleated nanobubbles can grow unlimitedly so that they can detach by buoyancy. {For example, by balancing the buoyancy force with the adhesion force from the electrode \citep{oguz1993dynamics}, the minimum detachment radius (`Fritz radius') of the bubble $R_d=\left(\frac{3\gamma L}{4\rho g}\right)^{1/3}$ ($\rho$ is the water density and $g$ is the gravity) is found to be approximately 3 $\mu$m.} {Similar to water electrolysis where the formation of gas bubbles on the electrode blocks the electrode and reduces the electrolysis efficiency, the formation of vapor bubbles on the heating solid during boiling also reduces the heat transfer efficiency. Thus a knowledge of self-rewetting techniques in boiling \citep{abe2006self,dhillon2015critical} may be helpful in improving the detachment of unstable nanobubbles in water electrolysis.} In our MD simulations, the threshold gas flux is found to be between $10.2$ kg/(m\textsuperscript{2}s) and $12$ kg/(m\textsuperscript{2}s). The specific value of course depends on the specifics of the system but the existence of a threshold value holds more generally. The observed features of nanobubbles in the diffusion-controlled growth are also very different from the macrobubbles studied before \cite{van2017electrolysis} which only consider the power-law growth of unstable bubbles, without the the possibility of stable bubbles and their transition to unstable bubbles.

\subsection*{Extended Lohse-Zhang model}
\begin{figure*}
\includegraphics[width=\linewidth]{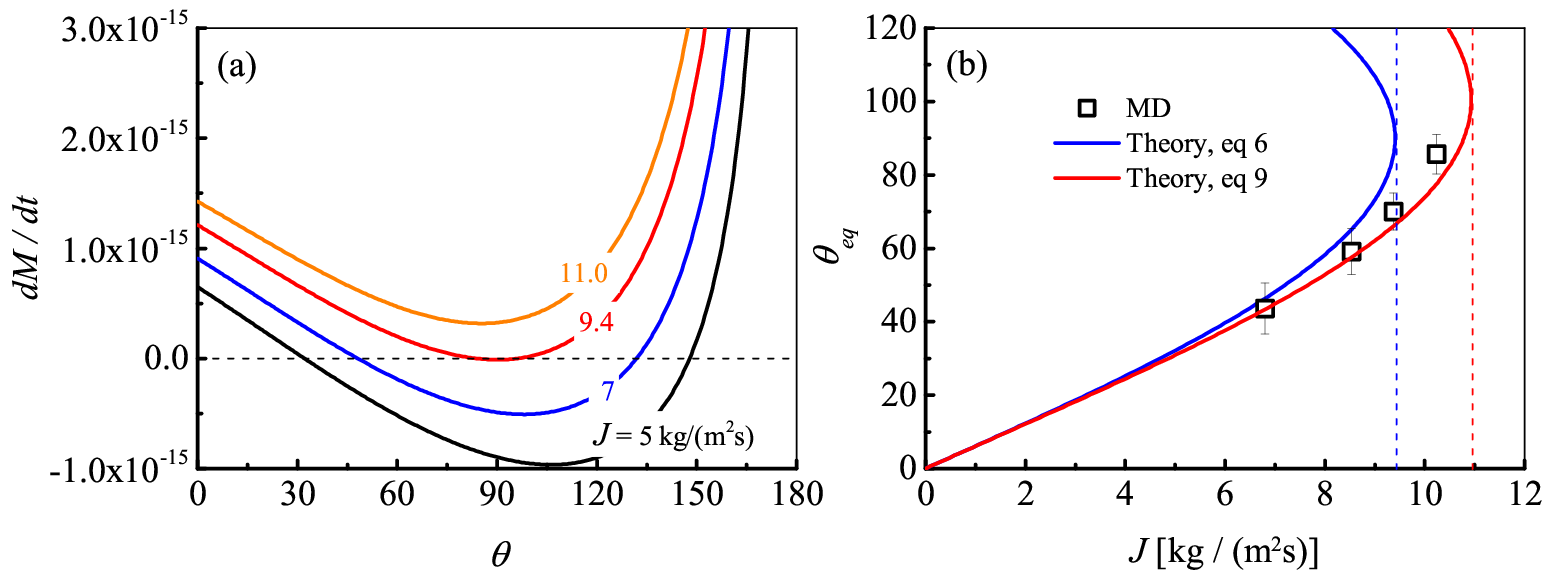}
\caption{(a) The mass change rate as a function of contact angle for four different gas fluxes. (b) The equilibrium contact angles as a function of gas fluxes. A comparison is made between MD simulations and theories. The error bars of $\theta_{eq}$ represent the standard deviation of 100 times measurements of contact angles performed in last 20 ns of simulations, see SI for more details.}\label{fig3}
\end{figure*}
In this section, we explain our MD results in the framework of the generalized Lohse-Zhang model \cite{lohse2015pinning,zhang2023minimum}. The growth of macroscopic bubbles on electrodes in water electrolysis is known to have two main modes, depending on the values of Damk\"{o}hler number $\text{Da}=A_e/R^2$, which is the ratio of active electrode area to the bubble surface area \cite{van2017electrolysis}. Note that this definition is traditionally used for spherical bubbles with contact angles $\theta=180^\circ$. For surface bubbles with a small contact angle but a very large radius $R$, this definition is obviously incorrect. Thus for bubbles with $\theta \le 90^\circ$, it may be better to use $R=L/2$, where $L$ is the pinning diameter of the bubble. For hemispherical bubbles in our simulations we find $\text{Da} \approx 36$. As will be seen later, stable nanobubbles in our simulations have contact angles of at most about $90^\circ $. Therefore the growth of nanobubbles in our current simulations is diffusion-controlled.

{The change of the bubble's mass $M$ depends on the rate-limiting step, which can either be the transfer rate of the gas across the liquid-gas interface or the rate of gas diffusing around the bubble. As here the latter is the case, because the time scale for the gas transport through the bubble surface is only about 0.1 ns \citep{popov2005evaporative}), much smaller than the diffusive time scale $L^2/D$ (which is about 19.3 ns in this work), we can assume a quasi-stationary diffusion equation and analytically solve it with the boundary conditions $C=C_R$ at the bubble surface and $C=C_{\infty}$ at the far field. This leads to the change rate of bubble mass \cite{lohse2015pinning}:}
\begin{equation}\label{eq:mass_rate}
\frac{dM}{dt}=-\frac{\pi }{2}LD\left( C_R -C_{\infty} \right)f\left( \theta  \right),\,\,\,\mbox{with}
\end{equation}
\begin{equation}
f\left( \theta  \right)=\frac{\sin \theta }{1+\cos \theta }+4\int_{0}^{\infty }{\frac{1+\cosh 2\theta \xi }{\sinh 2\pi \xi }}\tanh \left[ \left( \pi -\theta  \right)\xi  \right]d\xi.
\end{equation}
Here $M$ is the mass of the bubble. $C_{\infty}$ is the gas concentration at the far field. $C_R$ is the gas concentration on the bubble surface given by the Henry's law:
\begin{equation}
C_R=\left(P_{\infty}+\frac{4\gamma \sin{\theta}}{L}\right)\frac{C_s}{P_{\infty}}
\end{equation}
Combining eq 1 and eq 3 together, one obtains
\begin{equation}\label{eq:mass_rate2}
\frac{dM}{dt}=-\frac{\pi }{2}LD{{C}_{s}}\left( \frac{4\gamma }{L{{P}_{\infty}}}\sin \theta -\zeta  \right)f\left( \theta  \right),
\end{equation}
Here $\zeta=C_{\infty}/C_s-1$ is the gas oversaturation. To have a stable nanobubble, one must have $\zeta>0$. During the water splitting, the gas produced by the electrochemical reaction creates the local oversaturation so that $\zeta>0$. Obviously, one has to connect the oversaturation $\zeta$ with the gas flux $J$ to make predictions for what can be observed in the MD simulations.

In our simulations, we implement the constant flux boundary condition at $z=0$ and enforce a constant gas concentration condition at $z=H$. A linear concentration profile $C(z)=C_s+J\left(H-z\right)/D$ is expected to emerge at the steady state. 

Assuming that the cell height $H\gg h$, the produced (linear) gas concentration profile varies slowly along the bubble height, so that for the small heights, the local gas concentration around the bubble is approximate to be constant at the value of $C(z=0)=C_s+JH/D$. Thus the oversaturation around the bubble is simply given by 
\begin{equation}
\zeta=JH/\left( DC_s \right).
\end{equation}
The Lohse-Zhang model with eq 5 indeed predicts that a threshold gas flux exists, as shown in Figure \ref{fig3} (a). Below $J_t=9.4$ kg/(m\textsuperscript{2}s), the curve (e.g., $J=5$ kg/(m\textsuperscript{2}s)) intersects with the line of $dM/dt=0$ with two points where the first point with negative gradient is stable, denoting the equilibrium contact angle. Above $J_t=9.4$ kg/(m\textsuperscript{2}s), the curve (e.g., $J=11.0$ kg/(m\textsuperscript{2}s)) is above the line of $dM/dt=0$ and there are no intersection points so that the nucleated bubble for this flux is unstable. When $dM/dt=0$, the relation between the equilibrium contact angles and the gas fluxes is found to simply be:
\begin{equation}\label{eqeqa}
\sin\theta_{eq}=\frac{JLHP_{\infty}}{4DC_s\gamma}.
\end{equation}

In Figure \ref{fig3} (b), the estimated equilibrium contact angles from MD simulations (black squares) are then compared to eq \ref{eqeqa} (the blue line). It seems that eq \ref{eqeqa} overpredicts the contact angles of MD results. The threshold gas flux is expected to be
\begin{equation}\label{eq7}
J_t=\frac{4DC_s\gamma}{LHP_{\infty}}\mathrm{max}\left(\sin\theta\right)=\frac{4DC_s\gamma}{LHP_{\infty}},
\end{equation}
which gives $J_t=9.4$ kg/(m\textsuperscript{2}s) (evaluated at $\theta=90^\circ$). This value is below the threshold found in MD simulations. Such deviations are expected since the model works for large cells where $H\gg h$. But in our simulations, $h=2.88$ nm, which is not small enough compared to $H=11.25$ nm.  
\begin{figure*}
\includegraphics[width=\linewidth]{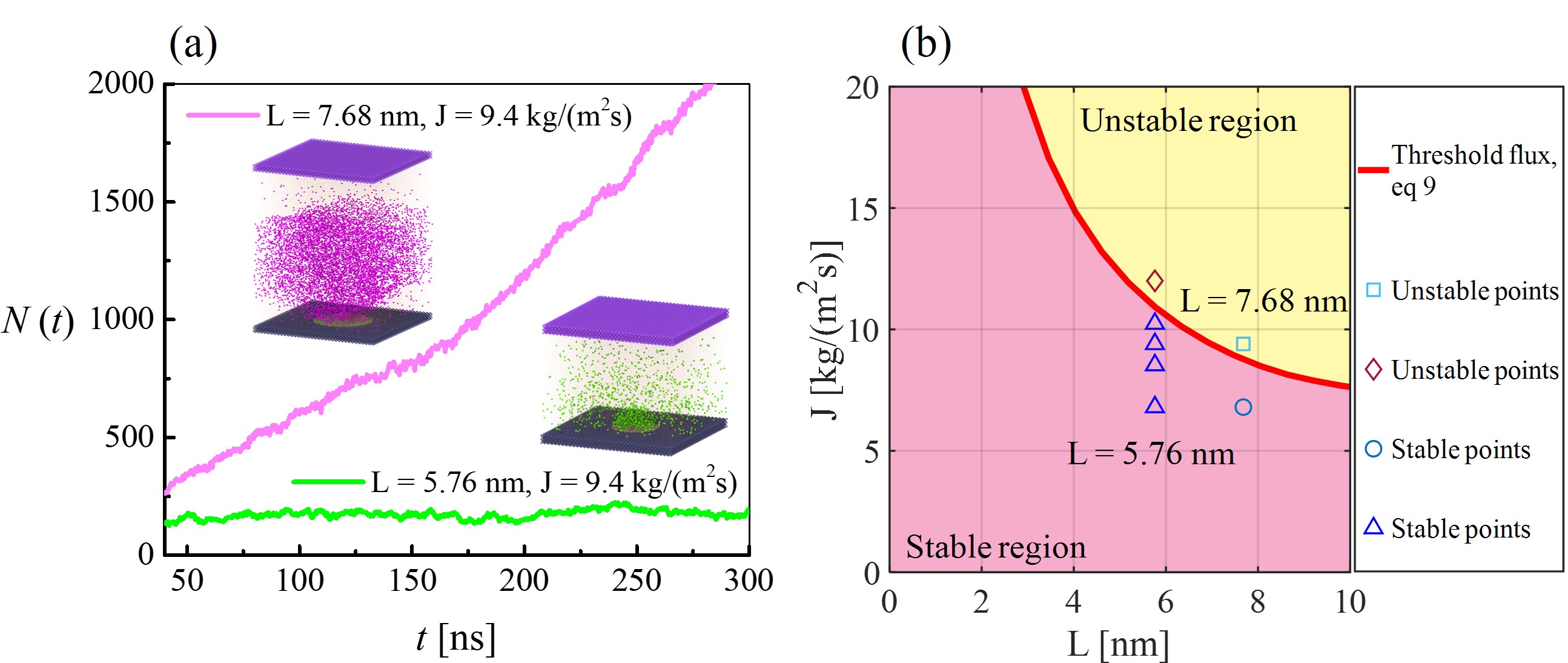}
\caption{(a) The evolution of the number of gas atoms inside bubbles for two cases with different pinning length $L=7.68$ nm and $L=5.76$ nm . The two snapshots show the bubble status for two cases respectively. For $L=5.76$ nm, bubble is stable while bubble is unstable for $L=7.68$ nm. (b) The phase diagram for stable and unstable bubbles. The red curve of the threshold gas flux is given by eq \ref{eqeqa2}.}
\label{fig4}
\end{figure*}
But we can correct for this: as the oversaturation profile next to the bubble is $\zeta(z)=J(H-z)/\left(DC_s\right)$, a simple correction to eq \ref{eqeqa} can be made by using the averaged oversaturation from $z=0$ to the bubble height $z=h$
\begin{equation}
\zeta=\frac{\int_0^h \zeta(z)dz}{h}=\frac{J}{Dc_s}\left(H-\frac{h}{2}\right)
\end{equation}
where $h=L(1-\cos\theta)/(2\sin\theta)$ based on the geometric definition. Then when $dM/dt=0$, the implicit equation for the equilibrium contact angle is
\begin{equation} \label{eqeqa2}
\sin\theta_{eq}=\frac{JLHP_{\infty}}{4DC_s\gamma}\left[1-\frac{L\left(1-\cos\theta_{eq}\right)}{4H\sin\theta_{eq} }\right].
\end{equation}
The numerical solution to eq \ref{eqeqa2} (see the red line) in Figure \ref{fig3} agrees excellently with MD simulations. The threshold current density is found to be 11 kg/(m\textsuperscript{2}s), when evaluated at the contact angle $100^{\circ}$, also in agreement with the MD simulations (between $10.2$ kg/(m\textsuperscript{2}s) and $12$ kg/(m\textsuperscript{2}s)). {In the current MD simulations, the simulated gas has a larger solubility compared to hydrogen. Based on eq \ref{eqeqa} and eq \ref{eq7}, the contact angle of a hydrogen bubble under a same current density will be larger and therefore the threshold current density will be smaller.}  

Notably, the constant value of $H=11.25$ nm is used in above calculations. This is indeed true in our simulations for small bubbles with contact angles up to about $90^{\circ}$ which happens to be the threshold contact angle for stable nanobubbles. If the threshold contact angle is larger, the usage of a constant $H$ will be problematic since the MD system will adjust $H$ significantly to maintain the ambient pressure. For real experiments, the condition of small bubble size in comparison to the cell size is easily satisfied so that there is no concerns about a constant $H$. Another issue in the MD simulations is that the initial \emph{diffusion-controlled} growth of bubbles will finally enter into the reaction-controlled regime as the bubble becomes very large compared to the electrode size. However, the transition from stable bubbles to unstable bubbles takes place when the bubble is still small (contact angles about $90^{\circ}$) and in the regime of diffusion-controlled growth.

\subsection*{The effects of the pinning length $L$}
For an electrode surface, chemical and
geometrical heterogeneities are usually unavoidable, which leads to contact line pinning and the variety of the pinning length. Here we investigate the effects of the pinning length on the bubble stability. 

The pinning length $L$ was increased from $L=5.76$ nm to $L=7.68$ nm. For the gas flux $J=9.4$ kg/(m\textsuperscript{2}s), the nucleated nanobubble with the pinning length $L=5.76$ nm is stable as shown by the snapshot in the lower right corner of Figure \ref{fig4} (a). The green line in Figure \ref{fig4} (a) proves that the number of gas atoms inside this bubble reaches equilibrium. For the same gas flux, however, the nucleated nanobubble becomes unstable when the pinning length is $L=7.68$ nm, which is revealed by the snapshot in the top left corner of Figure \ref{fig4} (a). Also the number of gas atoms inside this bubble (the pink line in Figure \ref{fig4} (a)) confirms that the bubble undergoes unlimited growth. 

Figure \ref{fig4} (b) shows the phase diagram for stable and unstable nanobubbles in the parameter space of gas flux $J$ and pinning length $L$. The red curve (eq \ref{eqeqa2}) shows that the threshold gas flux density decreases with the pinning length, which means that nanobubbles with larger pinning length become more unstable. The red line well predicts the behaviors of bubble stability observed in our MD simulations, see the denotation of various symbols in Figure \ref{fig4} (b).  
\subsection*{Continuum numerical method}
\begin{figure*}
\includegraphics[width=\linewidth]{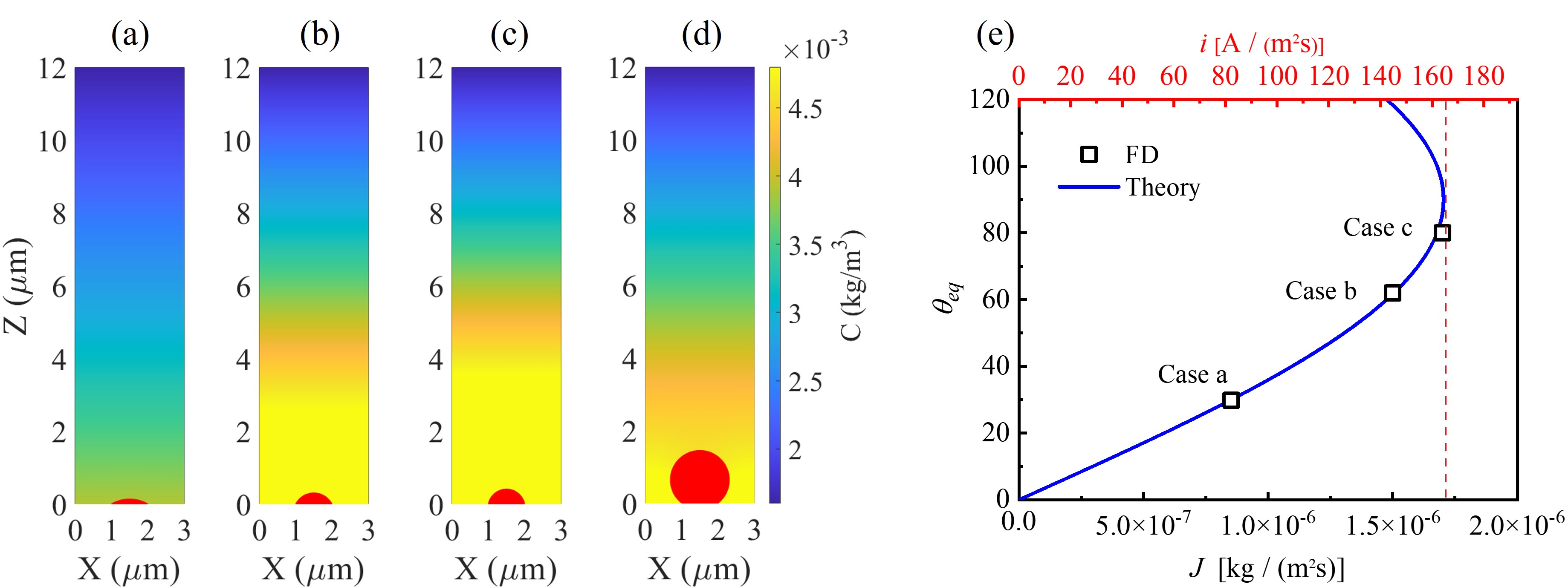}
\caption{The diffusion stability of bubbles simulated in FD under four different gas fluxes. (a) $J=0.85\times 10^{-6}$ kg/(m\textsuperscript{2}s), stable bubble; (b) $J=1.5\times 10^{-6}$ kg/(m\textsuperscript{2}s), stable bubble; (c) $J=1.7\times 10^{-6}$ kg/(m\textsuperscript{2}s), stable bubble; (d) $J=2.6\times 10^{-6}$ kg/(m\textsuperscript{2}s), unstable bubble where the bubble will eventually touch the side wall, which is nonphysical given the periodical boundary condition. Note that in this unstable case, the gas concentration profile $C(x,z,t)$ is not in an equilibrium state. (e) The relation between equilibrium contact angles and gas fluxes (current densities) for stable bubbles.}\label{fig5}
\end{figure*}
The condition $H \gg h$ is impractical to achieve in MD simulations due to the intensive computational costs. For a large system with $H \gg h$, we resort to the computationally more efficient finite difference (FD) method combined with the immersed boundary method (IBM). This method is described in Zhu et al  \cite{zhu2018diffusive}. The unsteady diffusion equation is solved with a constant gas concentration condition $C=C_s$ at the top boundary and a constant flux condition ${\partial C}/{\partial z}=J/D$ at the bottom boundary.

The parameters we choose are: the water surface tension 72 mN/m, hydrogen solubility $1.6\times 10^{-3}$ kg/m\textsuperscript{3}, and mass diffusivity $4.5\times 10^{-9}$ m\textsuperscript{2}/s \citep{cussler2009diffusion}. The pinning length of the bubble is 1 $\mu$m. The lateral length of the bottom electrode is 3 $\mu$m. The height of the system is 12 $\mu$m. 

By varying the gas flux $J$ (see the caption of Figure \ref{fig5}), we obtain different states of bubbles (see Movie S4 for the case with $J=1.5\times 10^{-6}$ kg/(m\textsuperscript{2}s)). For the cases (a-c) shown in Figure \ref{fig5}, bubbles are stable with different equilibrium contact angles, which are shown by the symbols in Figure \ref{fig5} (e). The computed equilibrium contact angles by FD agree well with the theory, i.e., eq \ref{eqeqa}. As the condition $H\gg h$ is satisfied, the theoretical description appears sufficient. For the case (d), the bubble is not stable (see Movie S5), which makes the system nonphysical when the bubble becomes so large that it touches the side boundary in conflict with the periodic boundary condition.
\subsection*{Comparison with experiments}
Currently, there are no experiments specifically addressing the diffusion-controlled stability of single nanobubbles on electrodes exposed to a current. Consequently, direct comparisons between experiments and the findings discussed in this study are not feasible. However, there are a number of experiments where multiple stable nanobubbles are produced on electrodes \cite{kikuchi2007characteristics,zhang2006electrochemically,yang2009electrolytically,yu2023interfacial}.  Assuming that these nanobubbles are isolated from each other (through sufficient distances between individual nanobubbles) allows for comparisons between these experiments and the theory presented in this work.

For example, Yang et al.  \cite{yang2009electrolytically} used an electrode of 10 mm $\times$ 10 mm and the cell size $H$ is estimated to be 10 mm. For the experiment operated under the cell potential of 1 V, the electrode is covered with multiple stable nanobubbles with a low density. The steady current is about 8 $\mu A$. Thus the current density is estimated to be 0.08 A/m\textsuperscript{2}, which translates a hydrogen flux of $8.3\times 10^{-9}$ kg/(m\textsuperscript{2}s). In the 1 V experiment, the typical pinning length $L$ of a stable nanobubble is about 200 nm and the bubble height is about 5 nm. Using the theory eq \ref{eqeqa}, the calculated contact angle of the stable nanobubble is about $4.6^\circ$, which is actually in good agreement with experimentally reported angle $5.7^\circ$.
\section*{Conclusions}\label{sec6}
In this work, the diffusion-controlled stability of single surface nanobubbles generated electrochemically is investigated by both numerical simulations (both MD and FD+IBM) and analytical theories. The current density is varied in simulations and it is found that nucleated nanobubbles can either be stable or unstable, depending on the value of the current density. This leads to the conclusion that there exists a threshold current density for stable nanobubbles in simulations. To theoretically explain this finding, the Lohse-Zhang model is extended by linking the current density with the local oversaturation, which nicely predicts the threshold current density found in the MD simulations. For stables nanobubbles, the theory also predicts equilibrium contact angles in agreement with simulations and experiments. By increasing the pinning length (which may be induced by coalescence of bubbles in practice), nanobubbles more easily become unstable since the threshold current density is reduced. This theory thus explains why some nanobubbles can adhere to electrode while others can become visible on a macroscale and then detach by buoyancy. The simulations and the theories presented here motivate new experiments to study electrolytic nanobubbles. Besides to water splitting, our conclusions should be applicable to other systems including electrochemical or catalytic gas evolution.
\subsection*{Supporting Information Appendix (SI)}
The legends of supporting movies and the method to obtain contact angles from MD simulations are described in SI Appendix.
%\subsubsection*{SI Movies}

%\begin{itemize}
  %\item SupportingInformation.pdf: simulation details.
  %\item Movie S1: Stable nanobubble in molecular simulations with $J=8.5$ kg/(m\textsuperscript{2}s).
  %\item Movie S2: Stable nanobubble in molecular simulations with $J=10.2$ kg/(m\textsuperscript{2}s).
  %\item Movie S3: Unstable nanobubble in molecular simulations with $J=12.0$ kg/(m\textsuperscript{2}s).
  %\item Movie S4: Stable microbubble in finite difference with $J=1.5\times 10^{-6}$ kg/(m\textsuperscript{2}s).
  %\item Movie S5: Unstable microbubble in finite difference with $J=2.6\times 10^{-6}$ kg/(m\textsuperscript{2}s).
%\end{itemize}

\matmethods{The mW water potential is adopted to model water. The mW water model is a monatomic water model proposed by Molinero and Moore\,\cite{molinero2009water}, and it uses the Stillinger-Weber (SW) potential whose parameters can be found in Ref \cite{molinero2009water}. 

Except for water itself, the intermolecular potentials $U$ between $i$-type atoms and $j$-type atoms are simulated with the standard Lennard-Jones (LJ) 12-6 potential:
\begin{equation}
  U({{r}_{ij}}) =
    \begin{cases}
      4\varepsilon_{ij} \left[ {{\left( \frac{\sigma_{ij} }{{{r}_{ij}}} \right)}^{12}}-{{\left( \frac{\sigma_{ij} }{{{r}_{ij}}} \right)}^{6}} \right] & \text{if} \,\,\,{{r}_{ij}}\le {{r}_{c}},\\
      0 & \text{if}\,\,\, {{r}_{ij}}>{{r}_{c}},\\
    \end{cases}       
\end{equation}
where $r_{ij},\varepsilon_{ij},\sigma_{ij}$ and ${r}_{c}$ are the pairwise distance, energy parameter, length parameter, and cut-off distance, respectively. The cutoff distance is chosen as $r_c=1.65$ nm. The complete list of parameters among the water (W), gas (G), hydrophobic electrodes (E\textsubscript{o}), hydrophilic electrodes (E\textsubscript{i}), and piston (P) are given in Table \ref{tab1}. 
\begin{table}[h!]
\centering
\caption{{\label{tab1}}Interaction parameters.}
 \begin{tabular}{c c c c} 
 Atom type & Atom type  & $\varepsilon_{ij} $(kcal/mol) & $\sigma_{ij}$ (Angstroms)  \\ 
 \hline
  G & G & 0.188 & 3.75    \\ 
 \hline
  W &  G & 0.20 & 3.07   \\
 \hline
  W & E\textsubscript{o} & 0.15 & 3.32 \\
  \hline
  G & E\textsubscript{o} & 0.26 & 3.32      \\ 
  \hline
  W & E\textsubscript{i} & 0.8 & 3.32 \\ 
  \hline
  W & P & 0.5 & 3.32    \\ 
  \hline
  G & P & 0.5 & 3.32    \\ 
  \hline
\end{tabular}
\end{table}

Gas atoms have the energy parameter $\varepsilon=0.188$ kcal/mol and the distance parameter $\sigma=0.375$ nm, and have a density $\rho_{\infty}=11.47$ kg/m\textsuperscript{3} at 10 atm and 300 K. The molar mass of the gas is 28 g/mol. The gas solubility $c_s$ is calculated by simulating the coexistence of water and gas and $c_s=0.54$ kg/m\textsuperscript{3} for $\epsilon_{WG}=0.20$ kcal/mol is obtained. The mass diffusivity of gas in water can be calculated by the Einstein-Stokes relation\,\citep{skoulidas2005self}. As the slope for the concentration distribution is $J/D$ in our simulations of electrolytic bubbles, it is used to obtain $D=4.3 \times 10^{-9}$ m\textsuperscript{2}/s here.  

By adjusting the water-electrode interaction $\varepsilon_{we}$, the hydrophobic electrode has a water contact angle $120^{\circ}$. Conversely, the hydrophilic electrode has a contact angle $5^{\circ}$. 

The box has a fixed lateral size with $L_x=17.28$ nm and $L_y=17.28$ nm. The height of this box is adjusted to maintain the far-field pressure $P_{\infty}=10$ atm where periodic boundary conditions are applied in the other two directions. The initial thickness of water slab is $12.5$ nm with 124416 atoms. The thickness of the bottom solid is $0.96$ nm and has a fcc structure with a number density $0.0332/{\textup{~\AA}}^3$. Based on the Butter-Volmer kinetics\,\citep{white2008electrochemistry} for electrochemical reaction, the reaction is not restricted to the first layer of water atoms above the electrode. Here the reaction zone above the electrode has a thickness $0.66$ nm corresponding to about the thickness of two or three layers of water atoms.

The far-field gas concentration is maintained at the gas solubility by switching the identity of gas atoms back into the identity of water atoms. This process is performed in a box with a thickness of $1.25$ nm placed below the piston plate. This lead to $H=11.25$ nm. Note that this process is carried out only when the gas concentration in the box is larger than the gas solubility. 
}

\showmatmethods{} % Display the Materials and Methods section

\acknow{We wish to acknowledge the financial support by NWO under the project of ECCM KICKstart DE-NL 20002799.}

\showacknow{} % Display the acknowledgments section

%\bibsplit[32]
\bibliography{pnas_nb.bib}

\begin{thebibliography}{10}

\bibitem{shih2022water}
AJ Shih, et~al., Water electrolysis.
\newblock {\em\protect\JournalTitle{Nat. Rev. Methods Primers}} \textbf{2}, 84
  (2022).

\bibitem{brauns2020alkaline}
J Brauns, T Turek, Alkaline water electrolysis powered by renewable energy: A
  review.
\newblock {\em\protect\JournalTitle{Processes}} \textbf{8}, 248 (2020).

\bibitem{yue2021hydrogen}
M Yue, et~al., Hydrogen energy systems: A critical review of technologies,
  applications, trends and challenges.
\newblock {\em\protect\JournalTitle{Renew. Sust. Energ. Rev.}} \textbf{146},
  111180 (2021).

\bibitem{ramachandran1998overview}
R Ramachandran, RK Menon, An overview of industrial uses of hydrogen.
\newblock {\em\protect\JournalTitle{Int. J. Hydrog. Energy}} \textbf{23},
  593--598 (1998).

\bibitem{vogt2005bubble}
H Vogt, R Balzer, The bubble coverage of gas-evolving electrodes in stagnant
  electrolytes.
\newblock {\em\protect\JournalTitle{Electrochim. Acta}} \textbf{50}, 2073--2079
  (2005).

\bibitem{angulo2020influence}
A Angulo, P van~der Linde, H Gardeniers, M Modestino, DF Rivas, Influence of
  bubbles on the energy conversion efficiency of electrochemical reactors.
\newblock {\em\protect\JournalTitle{Joule}} \textbf{4}, 555--579 (2020).

\bibitem{zhao2019gas}
X Zhao, H Ren, L Luo, Gas bubbles in electrochemical gas evolution reactions.
\newblock {\em\protect\JournalTitle{Langmuir}} \textbf{35}, 5392--5408 (2019).

\bibitem{karlsson2016selectivity}
RK Karlsson, A Cornell, Selectivity between oxygen and chlorine evolution in
  the chlor-alkali and chlorate processes.
\newblock {\em\protect\JournalTitle{Chem. Rev.}} \textbf{116}, 2982--3028
  (2016).

\bibitem{lu2015superaerophobic}
Z Lu, et~al., Superaerophobic electrodes for direct hydrazine fuel cells.
\newblock {\em\protect\JournalTitle{Adv. Mater.}} \textbf{27}, 2361--2366
  (2015).

\bibitem{papachristodoulou1985bubble}
A Papachristodoulou, F Foulkes, J Smith, Bubble characteristics and aerosol
  formation in electrowinning cells.
\newblock {\em\protect\JournalTitle{J. Appl. Electrochem.}} \textbf{15},
  581--590 (1985).

\bibitem{lohse2015surface}
D Lohse, X Zhang, Surface nanobubbles and nanodroplets.
\newblock {\em\protect\JournalTitle{Rev. Mod. Phys}} \textbf{87}, 981 (2015).

\bibitem{yang2009electrolytically}
S Yang, et~al., Electrolytically generated nanobubbles on highly orientated
  pyrolytic graphite surfaces.
\newblock {\em\protect\JournalTitle{Langmuir}} \textbf{25}, 1466--1474 (2009).

\bibitem{wang2010boundary}
Y Wang, B Bhushan, Boundary slip and nanobubble study in micro/nanofluidics
  using atomic force microscopy.
\newblock {\em\protect\JournalTitle{Soft Matter}} \textbf{6}, 29--66 (2010).

\bibitem{zhao2013mechanical}
B Zhao, et~al., Mechanical mapping of nanobubbles by peakforce atomic force
  microscopy.
\newblock {\em\protect\JournalTitle{Soft Matter}} \textbf{9}, 8837--8843
  (2013).

\bibitem{yu2023interfacial}
J Yu, et~al., Interfacial nanobubbles’ growth at the initial stage of
  electrocatalytic hydrogen evolution.
\newblock {\em\protect\JournalTitle{Energy Environ. Sci.}} \textbf{16},
  2068--2079 (2023).

\bibitem{luo2013electrogeneration}
L Luo, HS White, Electrogeneration of single nanobubbles at sub-50-nm-radius
  platinum nanodisk electrodes.
\newblock {\em\protect\JournalTitle{Langmuir}} \textbf{29}, 11169--11175
  (2013).

\bibitem{liu2017dynamic}
Y Liu, MA Edwards, SR German, Q Chen, HS White, The dynamic steady state of an
  electrochemically generated nanobubble.
\newblock {\em\protect\JournalTitle{Langmuir}} \textbf{33}, 1845--1853 (2017).

\bibitem{chen2015electrochemical}
Q Chen, HS Wiedenroth, SR German, HS White, Electrochemical nucleation of
  stable n2 nanobubbles at pt nanoelectrodes.
\newblock {\em\protect\JournalTitle{J. Am. Chem. Soc.}} \textbf{137},
  12064--12069 (2015).

\bibitem{edwards2019voltammetric}
MA Edwards, HS White, H Ren, Voltammetric determination of the stochastic
  formation rate and geometry of individual h2, n2, and o2 bubble nuclei.
\newblock {\em\protect\JournalTitle{ACS Nano}} \textbf{13}, 6330--6340 (2019).

\bibitem{suvira2023imaging}
M Suvira, et~al., Imaging single h2 nanobubbles using off-axis dark-field
  microscopy.
\newblock {\em\protect\JournalTitle{Anal. Chem.}} \textbf{95}, 15893--15899
  (2023).

\bibitem{zhou2023nanopipettes}
H Zhou, et~al., Nanopipettes for single nanobubble electrochemical analysis:
  fundamentals and applications.
\newblock {\em\protect\JournalTitle{Curr. Opin. Electrochem.}} p. 101370
  (2023).

\bibitem{hao2018imaging}
R Hao, Y Fan, MD Howard, JC Vaughan, B Zhang, Imaging nanobubble nucleation and
  hydrogen spillover during electrocatalytic water splitting.
\newblock {\em\protect\JournalTitle{Proc. Natl. Acad. Sci.U.S.A.}}
  \textbf{115}, 5878--5883 (2018).

\bibitem{deng2022direct}
X Deng, et~al., Direct measuring of single--heterogeneous bubble nucleation
  mediated by surface topology.
\newblock {\em\protect\JournalTitle{Proc. Natl. Acad. Sci.U.S.A.}}
  \textbf{119}, e2205827119 (2022).

\bibitem{lemineur2021imaging}
JF Lemineur, et~al., Imaging and quantifying the formation of single
  nanobubbles at single platinum nanoparticles during the hydrogen evolution
  reaction.
\newblock {\em\protect\JournalTitle{ACS Nano}} \textbf{15}, 2643--2653 (2021).

\bibitem{perez2019mechanisms}
YA Perez~Sirkin, ED Gadea, DA Scherlis, V Molinero, Mechanisms of nucleation
  and stationary states of electrochemically generated nanobubbles.
\newblock {\em\protect\JournalTitle{J. Am. Chem. Soc.}} \textbf{141},
  10801--10811 (2019).

\bibitem{maheshwari2020nucleation}
S Maheshwari, C Van~Kruijsdijk, S Sanyal, AD Harvey, Nucleation and growth of a
  nanobubble on rough surfaces.
\newblock {\em\protect\JournalTitle{Langmuir}} \textbf{36}, 4108--4115 (2020).

\bibitem{ma2021dynamic}
Y Ma, Z Guo, Q Chen, X Zhang, Dynamic equilibrium model for surface nanobubbles
  in electrochemistry.
\newblock {\em\protect\JournalTitle{Langmuir}} \textbf{37}, 2771--2779 (2021).

\bibitem{gadea2020electrochemically}
ED Gadea, YA Perez~Sirkin, V Molinero, DA Scherlis, Electrochemically generated
  nanobubbles: invariance of the current with respect to electrode size and
  potential.
\newblock {\em\protect\JournalTitle{J. Phys. Chem. Lett.}} \textbf{11},
  6573--6579 (2020).

\bibitem{zhang2023minimum}
Y Zhang, D Lohse, Minimum current for detachment of electrolytic bubbles.
\newblock {\em\protect\JournalTitle{J. Fluid Mech.}} \textbf{975}, R3 (2023).

\bibitem{parker1994bubbles}
JL Parker, PM Claesson, P Attard, Bubbles, cavities, and the long-ranged
  attraction between hydrophobic surfaces.
\newblock {\em\protect\JournalTitle{J. Phys. Chem.}} \textbf{98}, 8468--8480
  (1994).

\bibitem{zhang2006physical}
XH Zhang, N Maeda, VS Craig, Physical properties of nanobubbles on hydrophobic
  surfaces in water and aqueous solutions.
\newblock {\em\protect\JournalTitle{Langmuir}} \textbf{22}, 5025--5035 (2006).

\bibitem{lou2000nanobubbles}
ST Lou, et~al., Nanobubbles on solid surface imaged by atomic force microscopy.
\newblock {\em\protect\JournalTitle{J. Vac. Sci. Technol.}} \textbf{18},
  2573--2575 (2000).

\bibitem{epstein1950stability}
PS Epstein, MS Plesset, On the stability of gas bubbles in liquid-gas
  solutions.
\newblock {\em\protect\JournalTitle{J. Chem. Phys.}} \textbf{18}, 1505--1509
  (1950).

\bibitem{lohse2015pinning}
D Lohse, X Zhang, Pinning and gas oversaturation imply stable single surface
  nanobubbles.
\newblock {\em\protect\JournalTitle{Phys. Rev. E}} \textbf{91}, 031003(R)
  (2015).

\bibitem{van2017electrolysis}
P Van Der~Linde, et~al., Electrolysis-driven and pressure-controlled diffusive
  growth of successive bubbles on microstructured surfaces.
\newblock {\em\protect\JournalTitle{Langmuir}} \textbf{33}, 12873--12886
  (2017).

\bibitem{higuera2021model}
F Higuera, A model of the growth of hydrogen bubbles in the electrolysis of
  water.
\newblock {\em\protect\JournalTitle{J. Fluid. Mech}} \textbf{927}, A33 (2021).

\bibitem{wang2016investigations}
Y Wang, X Hu, Z Cao, L Guo, Investigations on bubble growth mechanism during
  photoelectrochemical and electrochemical conversions.
\newblock {\em\protect\JournalTitle{Colloids Surf. A: Physicochem. Eng.}}
  \textbf{505}, 86--92 (2016).

\bibitem{verhaart1980growth}
H Verhaart, R De~Jonge, S Van~Stralen, Growth rate of a gas bubble during
  electrolysis in supersaturated liquid.
\newblock {\em\protect\JournalTitle{Int. J. Heat Mass Transf.}} \textbf{23},
  293--299 (1980).

\bibitem{lohse2016homogeneous}
D Lohse, A Prosperetti, Homogeneous nucleation: Patching the way from the
  macroscopic to the nanoscopic description.
\newblock {\em\protect\JournalTitle{Proc. Natl. Acad. Sci.U.S.A.}}
  \textbf{113}, 13549--13550 (2016).

\bibitem{jones1999bubble}
S Jones, G Evans, K Galvin, Bubble nucleation from gas cavities—a review.
\newblock {\em\protect\JournalTitle{Adv. Colloid Interface Sci.}} \textbf{80},
  27--50 (1999).

\bibitem{plimpton1995fast}
S Plimpton, Fast parallel algorithms for short-range molecular dynamics.
\newblock {\em\protect\JournalTitle{J. Comput. Phys.}} \textbf{117}, 1--19
  (1995).

\bibitem{dockar2018mechanical}
D Dockar, MK Borg, JM Reese, Mechanical stability of surface nanobubbles.
\newblock {\em\protect\JournalTitle{Langmuir}} \textbf{35}, 9325--9333 (2018).

\bibitem{molinero2009water}
V Molinero, EB Moore, Water modeled as an intermediate element between carbon
  and silicon.
\newblock {\em\protect\JournalTitle{J. Phys. Chem. B}} \textbf{113}, 4008--4016
  (2009).

\bibitem{bard2022electrochemical}
AJ Bard, LR Faulkner, HS White, {\em Electrochemical methods: fundamentals and
  applications}.
\newblock (John Wiley \& Sons), (2022).

\bibitem{penas2019decoupling}
P Pe{\~n}as, et~al., Decoupling gas evolution from water-splitting electrodes.
\newblock {\em\protect\JournalTitle{J. Electrochem. Soc.}} \textbf{166}, H769
  (2019).

\bibitem{brussieux2011controlled}
C Brussieux, P Viers, H Roustan, M Rakib, Controlled electrochemical gas bubble
  release from electrodes entirely and partially covered with hydrophobic
  materials.
\newblock {\em\protect\JournalTitle{Electrochim. Acta}} \textbf{56}, 7194--7201
  (2011).

\bibitem{zhang2019molecular}
Y Zhang, JE Sprittles, DA Lockerby, Molecular simulation of thin liquid films:
  Thermal fluctuations and instability.
\newblock {\em\protect\JournalTitle{Phys. Rev. E}} \textbf{100}, 023108 (2019).

\bibitem{weijs2011origin}
JH Weijs, A Marchand, B Andreotti, D Lohse, JH Snoeijer, Origin of line tension
  for a lennard-jones nanodroplet.
\newblock {\em\protect\JournalTitle{Phys. Fluids}} \textbf{23}, 022001 (2011).

\bibitem{rohsenow1971boiling}
WM Rohsenow, Boiling.
\newblock {\em\protect\JournalTitle{Annu. Rev. Fluid Mech.}} \textbf{3},
  211--236 (1971).

\bibitem{prosperetti2017vapor}
A Prosperetti, Vapor bubbles.
\newblock {\em\protect\JournalTitle{Annu. Rev. Fluid Mech.}} \textbf{49},
  221--248 (2017).

\bibitem{oguz1993dynamics}
HN Oguz, A Prosperetti, Dynamics of bubble growth and detachment from a needle.
\newblock {\em\protect\JournalTitle{J. Fluid Mech.}} \textbf{257}, 111--145
  (1993).

\bibitem{abe2006self}
Y Abe, Self-rewetting fluids: Beneficial aqueous solutions.
\newblock {\em\protect\JournalTitle{Ann. N. Y. Acad. Sci.}} \textbf{1077},
  650--667 (2006).

\bibitem{dhillon2015critical}
NS Dhillon, J Buongiorno, KK Varanasi, Critical heat flux maxima during boiling
  crisis on textured surfaces.
\newblock {\em\protect\JournalTitle{Nat. Commun.}} \textbf{6}, 8247 (2015).

\bibitem{popov2005evaporative}
YO Popov, Evaporative deposition patterns: spatial dimensions of the deposit.
\newblock {\em\protect\JournalTitle{Phys. Rev. E}} \textbf{71}, 036313 (2005).

\bibitem{zhu2018diffusive}
X Zhu, R Verzicco, X Zhang, D Lohse, Diffusive interaction of multiple surface
  nanobubbles: shrinkage, growth, and coarsening.
\newblock {\em\protect\JournalTitle{Soft Matter}} \textbf{14}, 2006--2014
  (2018).

\bibitem{cussler2009diffusion}
EL Cussler, {\em Diffusion: mass transfer in fluid systems}.
\newblock (Cambridge university press), (2009).

\bibitem{kikuchi2007characteristics}
K Kikuchi, S Nagata, Y Tanaka, Y Saihara, Z Ogumi, Characteristics of hydrogen
  nanobubbles in solutions obtained with water electrolysis.
\newblock {\em\protect\JournalTitle{J. Electroanal. Chem.}} \textbf{600},
  303--310 (2007).

\bibitem{zhang2006electrochemically}
L Zhang, et~al., Electrochemically controlled formation and growth of hydrogen
  nanobubbles.
\newblock {\em\protect\JournalTitle{Langmuir}} \textbf{22}, 8109--8113 (2006).

\bibitem{skoulidas2005self}
AI Skoulidas, DS Sholl, Self-diffusion and transport diffusion of light gases
  in metal-organic framework materials assessed using molecular dynamics
  simulations.
\newblock {\em\protect\JournalTitle{J. Phys. Chem. B}} \textbf{109},
  15760--15768 (2005).

\bibitem{white2008electrochemistry}
RJ White, HS White, Electrochemistry in nanometer-wide electrochemical cells.
\newblock {\em\protect\JournalTitle{Langmuir}} \textbf{24}, 2850--2855 (2008).

\end{thebibliography}

\end{document}